# Micro-Bunched Beam Production at FAST for Narrow Band THz Generation Using a Slit-Mask

J. Hyun†, SOKENDAI, Tsukuba, Ibaraki 305-0801, Japan
D. Crawford, D. Edstrom Jr, J. Ruan, J. Santucci, R. Thurman-Keup, T. Sen and J.C. Thangaraj,
Fermi National Accelerator Laboratory*, Batavia, Illinois 60510, USA

*Abstract*

We discuss simulations and experiments on creating micro-bunch beams for generating narrow band THz radiation at the Fermilab Accelerator Science and Technology (FAST) facility. The low-energy electron beamline at FAST consists of a photoinjector-based RF gun, two L-band superconducting accelerating cavities, a chicane, and a beam dump. The electron bunches are lengthened with cavity phases set off-crest for better longitudinal separation and then micro-bunched with a slit-mask installed in the chicane. We carried out the experiments with 30 MeV electron beams and detected signals of the micro-bunching using a skew quadrupole magnet in the chicane. In this paper, the details of micro-bunch beam production, the detection of micro-bunching and comparison with simulations are described.

## INTRODUCTION

Terahertz (THz) radiation, with a frequency between 0.1 THz - 10 THz, is non-ionizing and has high transmittance through clothes, paper, and plastic. Therefore, THz radiation is used in a wide range of fields such as medical treatment, homeland security, and material science. To be applied in such fields, a compact and intense THz source with tunable frequency is desired. THz sources with a laser or an accelerator have recently been proposed and developed. Yields of THz radiation generated by a combination of a laser and a nonlinear crystal are limited by damage to the crystal when a high power laser is used. Also, CSR, Smith-Purcell radiation, and Transition radiation emitted with an accelerator are spectrally broadband.

Our goal is to generate narrow band THz waves with a frequency of over 1 THz with a slit-mask and a linear accelerator based on superconducting structures. The advantage of this scheme is that intense THz waves can be generated due to the high repetition rate and high beam charge. Moreover, a tunable narrow-band THz wave can be produced using a slit-mask in a chicane and an energy-chirped electron beam.

We performed experiments on micro-bunch beam production for THz radiation at the Fermilab Accelerator Science and Technology (FAST) facility electron LINAC. We also observed micro-bunched electron beams using a skew quadrupole downstream of the slits in the chicane. Here we present simulations and experimental results on micro-bunched beam production and compare the measurements with simulations and analytical calculations.

† hyon@post.kek.jp

## FAST INJECTOR

In this section, the FAST photoinjector and parameters of the electron beam are described. Figure 1 and Table 1 show the layout and basic parameters of the photoinjector. This injector consists of a $Cs_2Te$ photocathode RF gun surrounded by two solenoid coils (bucking and main coils), two superconducting RF (SRF) TESLA-style 9-cell cavities, quadrupole magnets, a chicane, a vertical bending magnet, and a beam absorber [1]. The RF gun is identical to the one developed for the FLASH facility at DESY [2]. All RF cavities (RF gun, two capture cavities) are operated at an RF frequency of 1.3 GHz.

The electrons produced at the photocathode have an energy of ~5 MeV with an energy spread of ~0.1 % at the exit of the RF gun, and are then accelerated up to a maximum energy of 50 MeV through the SRF cavities. Transverse beam size is controlled with two triplets and a doublet of quadrupoles, and measured using YAG screens before and after the chicane.

A tungsten slit-mask with slit spacing, widths, and thickness of 950 μm, 50 μm, and 0.5 mm respectively (see Fig.1) is installed in the middle of the chicane. Given this ratio of the slit width to slit spacing, we expect about 5% of the beam to be cleanly transmitted through the slits, the remaining beam passing through the tungsten is scattered at large angles and lost at the beam pipe. This has been confirmed with Geant4 [3] simulations. Also, a skew quadruple is installed downstream of the slit-mask in the chicane to check for micro-bunch beam generation.

Table 1: FAST's injector parameters.

| Parameter | Value |
|---|---|
| Beam energy | <50 MeV |
| Bunch charge | 200 pC |
| Accelerating frequency | 1.3 GHz |
| Bunch frequency | 3 MHz |
| Normalized emittance | < 2 μm-rad |
| Energy spread | < 0.1% |
| Chicane dispersion | 0.18 m |
| Transfer matrix $R_{56}$ | 0.32 m |

## MICRO-BUNCHED BEAM PRODUCTION

We briefly explain our method of generating and observing a micro bunched beam in this section. When an electron beam is accelerated with an off-crest phase in cavities (see (a) & (b) in Fig. 1), the bunch has an energy spread $\Delta p/p$ ($=\delta$) correlated with its length $\sigma_z$, and the electron beam has an energy chirp (see (c) in Fig.1). The energy chirp $h$ can be computed from

$$h = \frac{d\delta}{d\sigma_z} \approx -\frac{eV_{rf}k_{rf}}{E}\sin\varphi_{rf}\,, \tag{1}$$

where $eV_{rf}$ is the energy gain, $\varphi_{rf}$ is the RF phase in the cavity, and $E$ is the beam energy after the last cavity. Here, $h<0$ means that high energy (low energy) electrons are at the bunch head (tail) after the cavity. The chirped electron beam is led to the chicane where there is horizontal dispersion and is separated in the horizontal plane (energy plane) using a slit-mask in the chicane (see (d) in Fig.1).

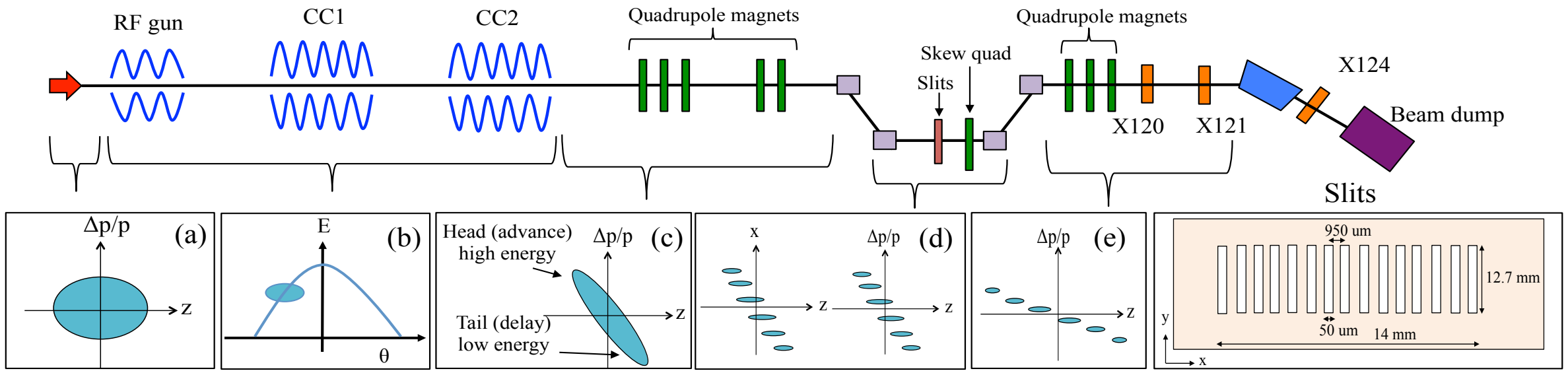

Figure 1: Layout of the FAST injector and phase spaces of an electron beam at each place.

After the chicane, the high-/low-energy electrons come to the bunch head/tail, therefore, the electron bunch is lengthened and a longitudinally separated micro-bunched beam is produced (see (e) in Fig.1). Then, the longitudinal width of the beamlets can be written as [4]

$$\sigma_{z,MB} = \frac{1}{|\eta h|}\sqrt{\eta^2\sigma_u^2 + \left(1+hR_{56}\right)^2\left[\frac{w}{2\sqrt{3}} + \varepsilon\beta\right]}\,, \tag{2}$$

where $\eta$ is the dispersion at the slit-mask, $h$ is the energy chirp, $w$ is the slit's width, $R_{56}$ is the transfer matrix of the chicane, $\varepsilon$ is the un-normalized transverse beam emittance, $\beta$ is the beta function at the slits and $\sigma_u$ is the uncorrelated energy spread. Eq. (2) indicates that $\varepsilon\beta$ and $\sigma_u$ should be small to create short beamlets with a clear separation among them. THz radiation is generated via transition radiation when the micro-bunched beam hits an Al foil downstream of the chicane. The lowest or fundamental frequency emitted is given by [4]

$$f_0 = \frac{\eta c|h|}{D|1+R_{56}h|}\,. \tag{3}$$

where $D$ is the slit spacing. Eq. (3) shows that the fundamental frequency can be tuned by changing the chirp.

To check the production of a micro-bunched beam, we use a skew quadrupole downstream of the slits in the chicane [5] (see Fig. 1). The skew quad creates a vertical dispersion and the vertical plane of the electron beam has information on separations in the horizontal plane. We can observe the vertical separations in the beam at a YAG screen after the chicane. We find that the average vertical spacing between the beamlets can be written as,

$$\langle \Delta y \rangle = \left(K_{sk}L\right)R_{34}D\,, \tag{4}$$

where $K_{sk}$ and $L$ are the K-value and length of the skew quadrupole, and $R_{34}$ is the transfer matrix element from the skew quadrupole to the beam monitor. The spacing in the vertical plane is proportional to the skew quad current but is independent of the chirp or RF phase.

## SIMULATIONS

This section presents simulation results on longitudinal distributions and frequency spectra. All simulations were performed using the “elegant” code [6].

Usable RF phases of the capture cavities CC1 and CC2 range from -35 degrees to 35 degrees. We chose the RF phases of (CC1, CC2) = (+30°, 0°) and (+30°, +30°) corresponding to energy chirps of $h$=-5.5 m$^{-1}$ and -12.5 m$^{-1}$, respectively. The accelerating voltages of the two cavities were tuned such that the final electron beam energy remained constant at ~30 MeV, and the beam optics was matched so that $\beta_x$ was 1 m at the slit-mask.

Figure 2 shows longitudinal distributions after the chicane for two energy chirps of -5.5 m$^{-1}$ and -12.5 m$^{-1}$. The widths are smaller for the larger energy chirp $|h|$, which is consistent with Eq. (2). Figure 3 shows the expected frequency spectra for the two energy chirps. The fundamental frequencies are 0.26 THz and 0.36 THz for $h$=-5.5 m$^{-1}$ and $h$=-12.5 m$^{-1}$, respectively, which agree with the analytical calculation from Eq. (3). Moreover, the highest significant peaks in the frequency spectrum are over 1 THz when $h$=-12.5 m$^{-1}$.

## EXPERIMENTS AT FAST

In this section, we show the experimental results on micro-bunched beam production. First, we focused the electron beam to a small size in the vertical plane at X120 (YAG screen) after the chicane because the vertical dispersion increases the vertical size when the skew quadrupole is turned on. The slit-mask in the chicane was then inserted and the skew quad set to 0.9 A. Figure 4 shows beam sizes at X120 measured (left plots) and simulated (right plots) at the RF phases of CC1=-29.4° (max. compression), 0° (min. energy spread), and +30° (max. lengthening). The beamlets are separated in the vertical-plane, and the spacing at the two off-crest phases (±30°) is about 0.8 mm. Eq. (4) shows that the spacing of the micro-bunched beam is independent of the chirp and therefore the measurements are in qualitative agreement with theory and simulations.

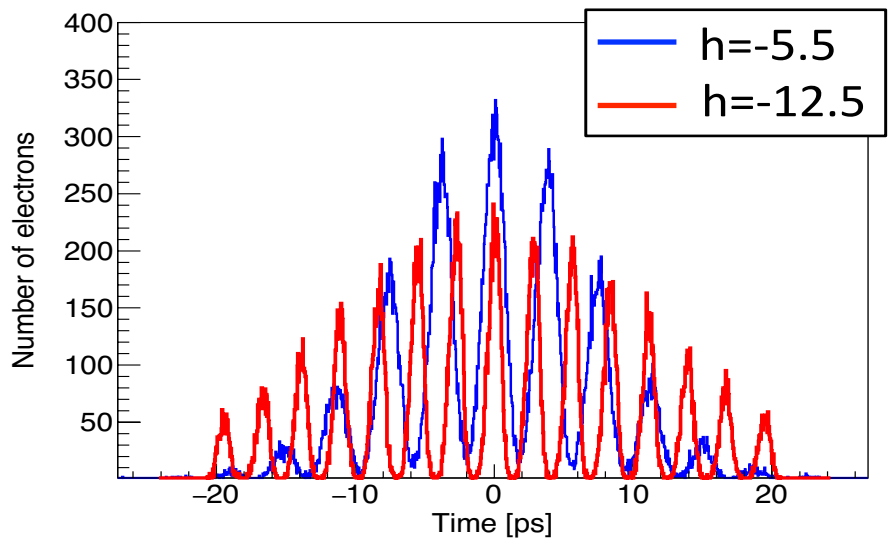

Figure 2: Longitudinal distributions after the chicane. Red and blue lines show h=-5.5 $m^{-1}$ and h=-12.5 $m^{-1}$, respectively.

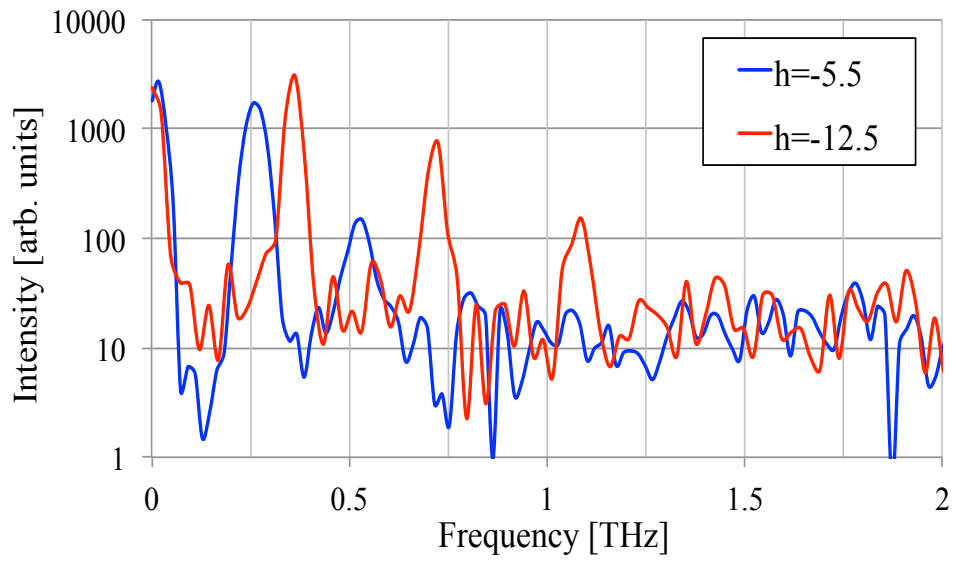

Figure 3: Frequency spectra for energy chirps of -5.5 $m^{-1}$ (blue line) and -12.5 $m^{-1}$ (red line).

Next, we set RF phases to (CC1, CC2) = (+30°, 0°) and to (0°, +30°) and scanned the current of the skew quadrupole. The results are shown in Figure 5. The vertical spacing is proportional to the skew quadrupole current, and there is no big difference between the two chirps, which is again consistent with Eq. (4).

Attempts were made to measure the radiation from the micro-bunched beam with a streak camera and a pyrometer. However, radiation signals of the micro-bunched beam could not be detected in the streak camera due to the low charge after the slit-mask and in the pyrometer due to large background from bremsstrahlung radiation generated at the slit-mask.

## CONCLUSIONS & OUTLOOK

Our simulations showed that radiation with frequencies in the THz range can be emitted when electrons are accelerated with sufficiently large off-crest phases in the SRF cavities. In the experiments we generated a micro-bunch beam and observed it by transforming the separation into the vertical plane using a skew quadrupole in the chicane. Moreover, measurements were in agreement with simulations and theory. During the next set of experiments, we will use a cryogenically cooled bolometer to monitor for THz radiation [7]. Background will be mitigated by a shielding system before the bolometer.

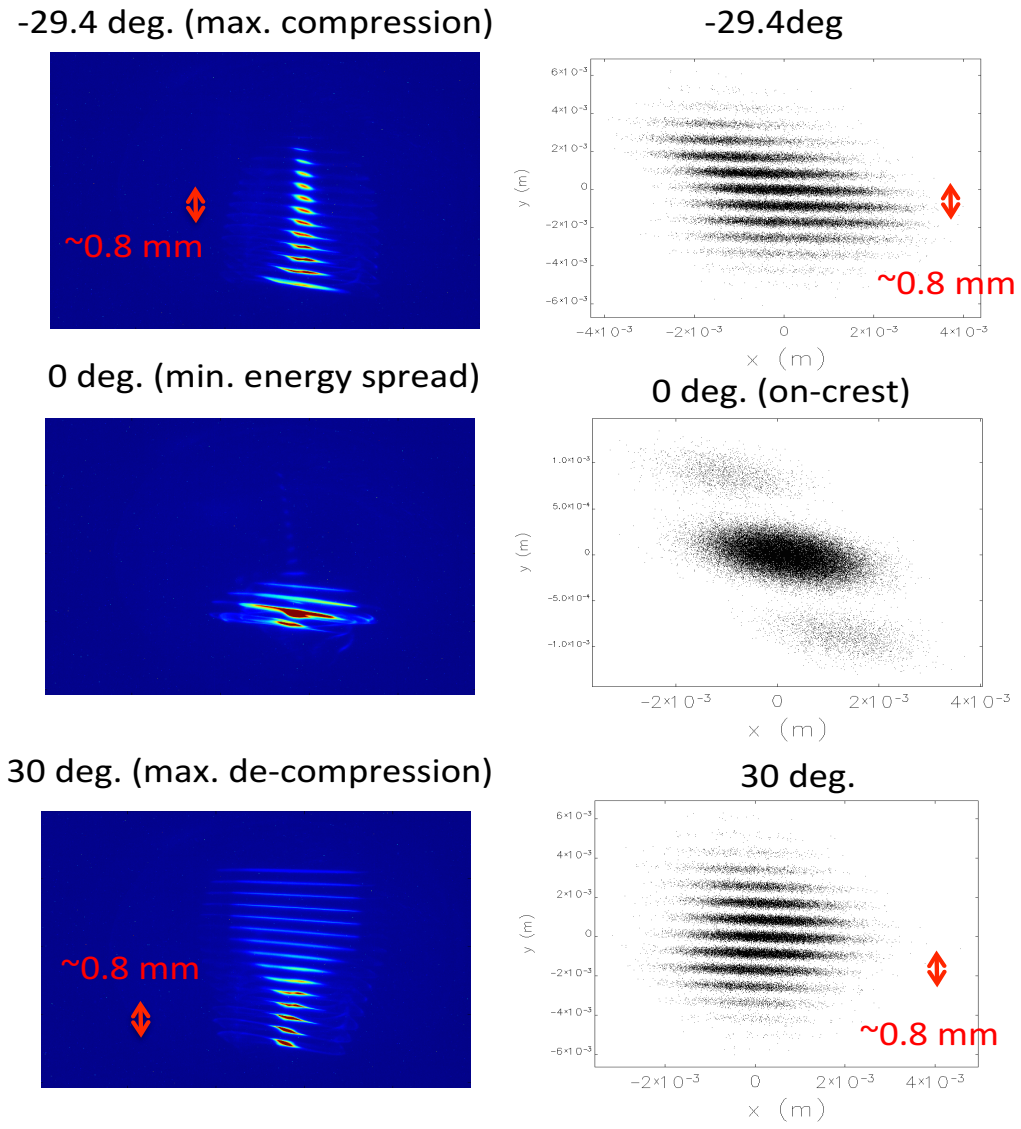

Figure 4: Beam distributions at X120 at CC1 RF phase of -29.4, 0, 30 degrees when the skew quad is set to 0.9 A. Left and right plots show measurements and simulations, respectively.

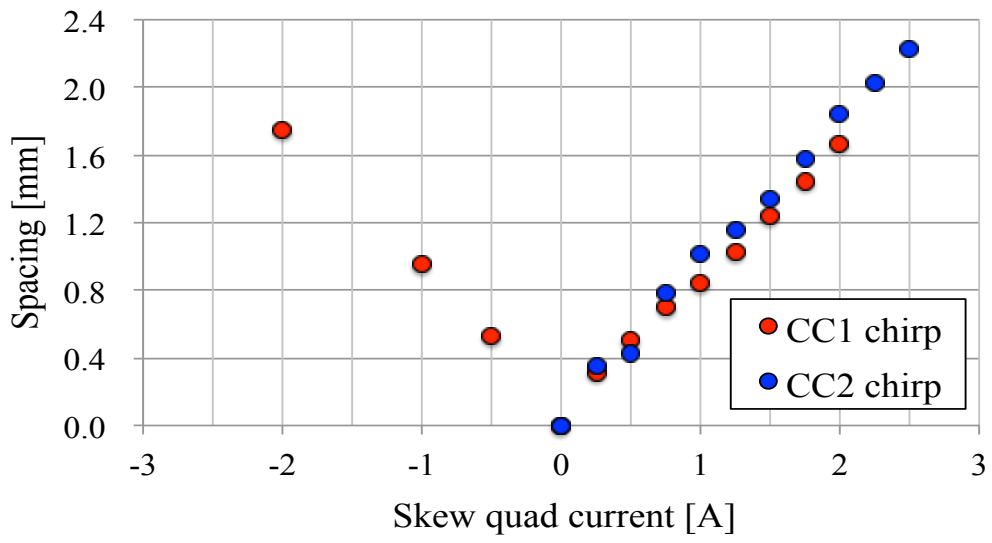

Figure 5: Vertical spacing depending on the skew quadrupole current at X120 at the off crest RF phases of CC1 (red dots) and CC2 (blue dots).

## REFERENCES

[1] D. Crawford, *et al.*, “First beam and high gradient cryomodule commissioning results of the advanced superconducting test accelerator at fermilab, in: Proceedings of IPAC2015, Richmond, VA, USA, 2015, p. 1831.

[2] B. Dwersteg, *et al.*, “RF gun design for the TESLA VUV Free Electron Laser”, Nucl. Instrum. Methods Phys. Res., Sect. A 393.

[3] Geant4, http://geant4.cern.ch.

[4] J. Thangaraj and P. Piot, “THz-radiation production using dispersively-selected flat beam bunches”, arXiv preprint arXiv:1310.5389v1 (2013)

[5] K. Bertsche, *et al.*, “A Simple, Low Cost Longitudinal Phase Space Diagnostic”, SLAC-PUB-13614 (2009).

[6] M. Borland, ”elegant: A Flexible SDDS-Compliant Code for Accelerator Simulation,” Advanced Photon Source LS-287, September 2000.

[7] P. Piot, et al., “Observation of coherently enhanced tunable narrow-band terahertz transition radiation from a relativistic sub-picosecond electron bunch train”, Appl. Phys. Lett. 98, 261501 (2011).